# The study of light nuclei production in different interactions at 4.2 AGeV/c

Kamal Hussain Khan,[1] M.K Suleymanov,[1,2] M. Ajaz,[3] Ali Zaman[1]

[1]Department of Physics, COMSATS Institute of Information Technology, Islamabad, Pakistan

[2]Joint Institute for Nuclear Research (JINR) Dubna, Russia

[3]Department of Physics, Abdul Wali Khan University Mardan, Pakistan

**Abstract**

Average multiplicity of light nuclei produced in different interactions at 4.2A GeV/c is studied as a function of centrality. A change in multiplicity is observed with increase in the mass of projectile. In $^{12}CC$-interactions an unexpected increase in the multiplicity is seen for the most central events. These measurements are compared with the predictions of Cascade and Fritiof Models, which fail to account for the experimentally observed effects. In case of $^{12}CC$, it is suggested that the inclusion of nuclear coalescence effect can be an explanatory reason for the differences between the experimental measurements and the models' predictions.

**Introduction**

In heavy ion collisions light nuclei are produced mainly from the fragmentation of projectile and target nuclei. However, nuclei are believed to be formed in final state interactions between nucleons as a result of coalescence [1], when they are in same phase space. Considering the momentum space, the probability density of nuclei formation of mass number A is proportional to the $A^{th}$ power of protons density [2-4]. A quantitative description of this process is typically based on the proportionality parameter $B_A$ known as coalescence parameter, which depends upon the transverse mass of cluster and does not depend upon the centrality of the collisions [5]. Whereas, the fragments of projectile and target decrease with the centrality of the collisions [6] and in the most central events, measurement of the yields of Deuterons and Tritons from target spectator area is suppressed [7]. At maximum centrality, the nuclei from either mechanism (fragmentation or coalescence) decrease exponentially with the mass number of produced nuclei [8]. So the study of the centrality dependent properties of light nuclei can give essential information about the initial and final states of collisions and the production mechanisms. In this paper the average multiplicity of light nuclei produced in Proton-Carbon ($p$C), Deutron-Carbon ($d$C), Helium-Carbon ($He$C) and Carbon-Carbon ($^{12}CC$)-interactions at 4.2 A GeV/c are studied as a function of centrality. The centrality is defined by the numbers of identified proton ($N_p$) in an event [9-11]. This study provides significant information about the behavior and production mechanisms of light nuclei.

**Experiment and Method**

The data were recorded with 2m Propane Bubble Chamber [12] at the Laboratory of High Energy of the Joint Institute for Nuclear Research (JINR), Dubna, Russia. The chamber was placed in a 1.5 T magnetic field, and irradiated with the beams of relativistic Protons (*p*), Deutron (*d*), Helium (*He*) and Carbon nuclei ($^{12}C$) with Propane ($C_3H_8$) as target. Almost all charged particles emitted at $4\pi$, provided they exceed the threshold value of the energy required for visible track formation were detected well in the chamber.

In total, we analyzed 12757 (twelve thousands seven hundred and fifty-seven) events of $pC_3H_8$, 9016 (nine thousands and sixteen) of $d\,C_3H_8$, 22975 (twenty-two thousands nine hundred and seventy-five) of $HeC_3H_8$ and 39543 (thirty-nine thousands five hundred and forty-three) of $^{12}C\,C_3H_8$. The interactions with Carbon nuclei were selected from all interactions of beams with Propane ($C_3H_8$), using criteria based on the determination of total charge of the secondaries, presence of slow Protons (with momentum P < 0.75 GeV/c), Protons in backward hemisphere and negatively charged particles etc. as described in Refs. [13,14]. This criterion is not used for *pC* interaction because light nuclei cannot be produced in *pp* interaction.

Particles were identified by their tracks which they left in the chamber and the momentum of the particles was calculated by the curvature of these tracks and the magnetic field of the chamber. The uncertainty in the measurement of momentum was about 11% and in measurement of emission angle θ, was about 0.8% [15,16].

Every particle required a minimum amount of momentum to produce a visible track particular to its mass. The average minimum momentum for Pion registration was set to about 70 MeV/c, below this Pion cannot produce visible tracks. All negative particles, except for those identified as electron were considered as $\pi^-$ mesons. The contamination from the misidentified electrons and negative strange particles were about 5% and 1% respectively. The $\pi^+$ mesons were identified and differentiated well from Protons by the ionization produced in the chamber in momentum region less than 0.5 GeV/c. Above this momentum $\pi^+$ mesons were mixed with Protons except a few which were recognized as $\pi^+$- mesons. The protons were identified finely within the momentum interval 0.15-0.5 GeV/c beyond this momentum Protons were contaminated with $\pi^+$ mesons. The nuclei were detected in two groups singly charged and multi-charged nuclei. The singly charged nuclei Deuteron (*d*) and Triton (*t*) were identified as mixture and differentiated well from other singly charged positive particles in the momentum interval 1-3GeV/c, further than this momentum and emission angle θ is less than 4° Deuteron and Triton could not be separated from striping Protons. The multi-charged nuclei with charge Z ≥ 2 were identified together. There was no possibility to identify the nuclei species separately because they generate about similar ionization in bubble chamber. Low momentum nuclei (multi or singly charged fragments) could not be identified due to their short length (invisible) tracks. In this manuscript all identified nuclei (singly or multi-charged) were considered as light nuclei. The centrality of the collisions was defined by the number of identified Protons ($N_p$) in an event, and identified Protons ($N_p$) was calculated as;

$N_p$= Protons (with any momentum) + $\pi^+$ mesons (with momentum >0.5 GeV/c) - $\pi^-$ mesons (with momentum >0.5 GeV/c) - Protons (with momentum >3 GeV/c and emission angle θ less than 4°).

As mentioned above the many of $\pi^+$ mesons with momentum greater than 0.5 GeV/c were identified as Protons, whereas $\pi^-$ mesons were identified very well. To address the contamination of $\pi^+$ mesons with Protons, it was assumed that equal number of $\pi^-$ and $\pi^+$ mesons were produced because the projectile (except proton) and target nuclei are Isospin singlet, that is why the $\pi^-$ mesons (with momentum >0.5 GeV/c) were subtracted from Protons. To deal with the contamination of *d* and *t*, the Protons with momentum greater than 3Gev/c and emission angle θ less than 4° were excluded (subtracted from Protons) also in defining $N_p$.

To observe the coalescence mechanism we used the simple idea of baryon number conservation and centrality. As the number of identified Protons in an event was used to fix the centrality of the collision, therefore an increase in the number of Protons (Increasing centrality) in an event will result in decrease in the number of nuclei (multiplicity of nuclei) to conserve the baryon numbers. So the study of multiplicity of light nuclei as a function of centrality can give some direct information about the production mechanism of nuclei in these collisions.

The experimental results were compared with the predictions of two theoretical models, Cascade [17 sec.3.1] and Fritiof [17 sec.3.2]. These codes are available on the website:

http:/hepweb.jinr.ru (created by V.V. Uzhinskii). Cascade model is used to describe the general features of relativistic nucleus-nucleus collisions. This model does not include any medium or collective properties and each of the colliding nuclei is treated as a gas of nucleons bound in a potential well. The Pauli principle and the energy momentum conservation are obeyed in each inter-nuclear interaction. The remaining excited nuclei, after the cascade stage, are described by the statistical theory in the evaporation approximation. Fritiof is a famous monte-carlo code which assumes all hadron-hadron interactions as binary reactions,($h_1+h_2 \rightarrow h'_1 + h'_2$), where $h'_1 + h'_2$ are excited states of hadrons with discrete or continues mass spectra. The excited hadrons are treated as QCD strings, and the corresponding Lund-string fragmentation model is applied in order to simulate their decays. The evaporation of residue nucleus is taken into account also. Similar approach is applied to simulate nucleus- nucleus collision, where successive interactions of projectile hadrons with target are considered. Unlike Cascade model, Fritof code uses the approach, which gives zero excitation energy to residual nucleus, when all spectator nucleon are ejected. The Cascade results can be reproduced by Fritiof model by changing limits of energy and distance between the nucleons. The Fritiof code has been modified [18] for lower energies and we used the modified version of this code. Both Models include nuclear fragments from projectile and target, but do not include the possibility of nuclei formation as a result of coalescence of nucleons. In both, the Cascade and the Fritiof codes we analyzed 40,000 (forty thousands) events of each interaction (*pC dC, HeC* and *$^{12}$CC*), under the same conditions and criteria, which was selected for experimental results for all the four interactions, using the same criteria which for experimental results.

**Results and Discussions**

The average multiplicity of the light nuclei ($<N>_{nuclei}$) in *p*C-interactions at 4.2A GeV/c is presented in Figure 1 as a function of centrality. The figure includes statistical uncertainties only. The measurements are compared with the predictions of two models, Cascade and Fritiof. Experimental results for average multiplicity are almost constant in *p*C -interactions. The observed light nuclei are the fragments of target nuclei (Carbon) only. More nuclei are identified in events with $N_p \geq 4$ and can be concluded that the target fragments (light nuclei) evaporated in these events are recorded relatively well. Both models' are predicting high values of multiplicities of light nuclei. As mentioned above the low momentum nuclei leave the invisible tracks in bubble chamber and cannot be recorded, which is the main reason of the diversity between the models' predictions and the experimental observation.

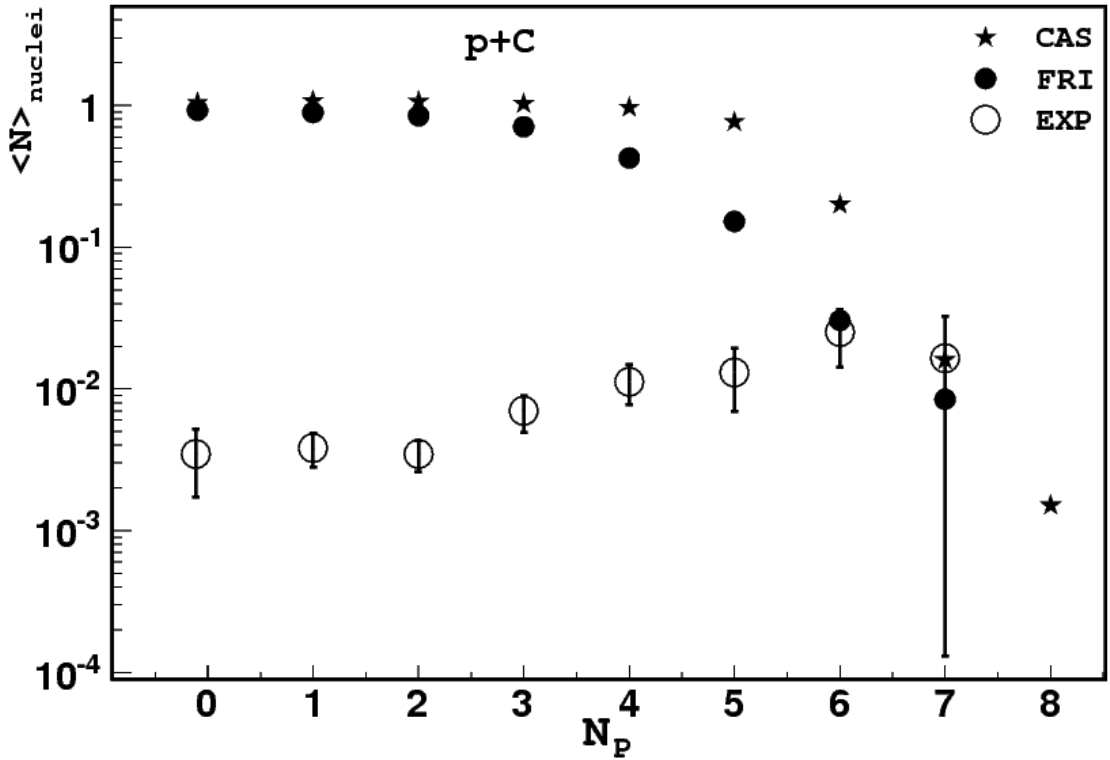

Figure 1. Average multiplicity of light nuclei as function of centrality ($N_p$) in $pC$ - interactions. Stars, solid circles and open circles represent Cascade (CAS), Fritiof (FRI) and data (EXP) respectively.

The average multiplicity of the light nuclei ($<N>_{nuclei}$) produced in $dC$-interactions at 4.2A GeV/c is presented in Figure 2 as a function of centrality. Only statistical errors are included in the figure. Now the projectile is itself a nucleus (Deuteron) and the total energy of the collision is double of the above interaction ($pC$-interactions), which increases the average multiplicity of light nuclei about 100 times greater than the multiplicity in $pC$- interactions. The experimental results of $dC$-interactions are almost constant except at first point ($N_p$=0), where the average multiplicity is maximum, which shows the contribution from projectile. The models' predictions are still deviating from the experimental results but lower as compared to the $pC$-interactions.

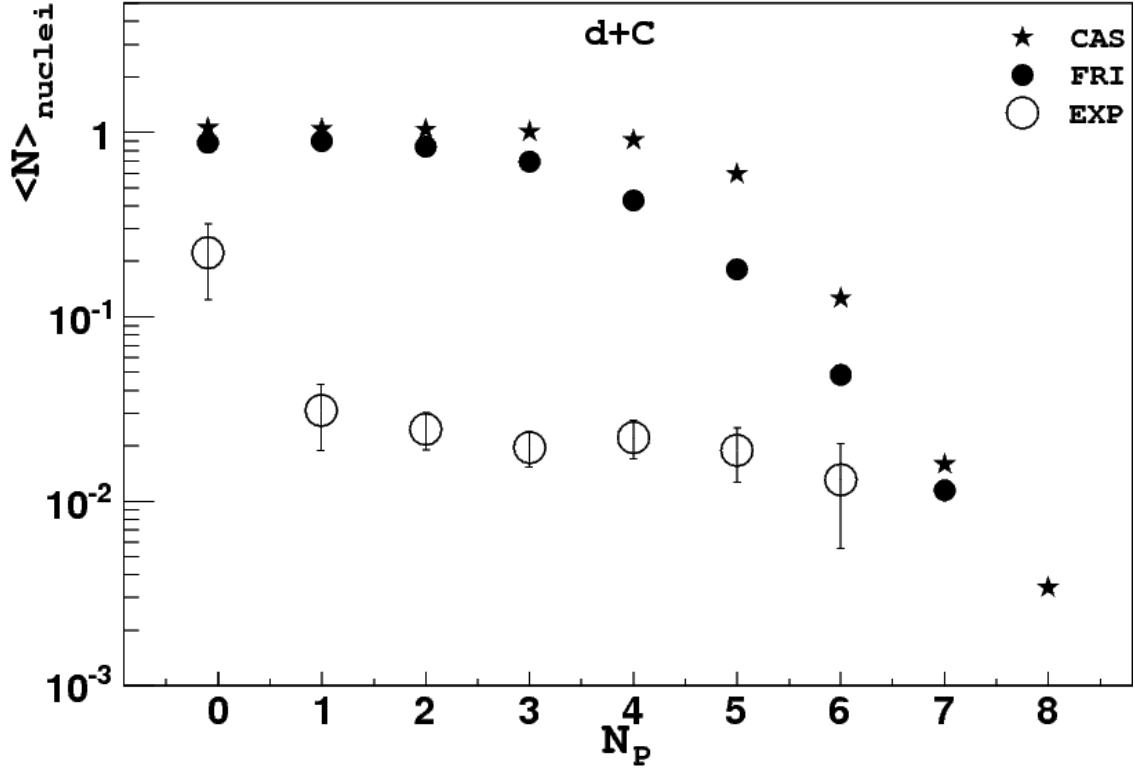

Figure 2. Average multiplicity of light nuclei as function of centrality ($N_p$) in $dC$-interactions. Stars, solid circles and open circles represent Cascade (CAS), Fritiof (FRI) and data (EXP) respectively

The average multiplicity of the light nuclei ($<N>_{nuclei}$) produced in $HeC$-interactions at 4.2A GeV/c is shown in Figure 3 as a function of centrality. Only statistical errors are considered in figure. Average multiplicity decreases from maximum to its minimum in region ($0 \leq N_p \leq 4$), and decreases slowly in region ($N_p > 4$) as shown above in $pC$ and $dC$-interactions. Both models are unable to reproduce the experimental results for average multiplicity of light nuclei; however the divergence between models' predictions and experimental results becomes small as compared to above mentioned interactions. In $HeC$-interactions the mass and energy of projectile is double of the $dC$-interaction, which increases the average multiplicity 10 times to $dC$-interactions and can be seen more clearly in figure 3. Keeping in view the above figures, the first region ($N_p < 4$) of figure 3 can be considered as projectile fragmenting region, whereas the second one ($N_p > 4$) is as target fragmenting region. The projectile fragmenting region is more sensitive for centrality and target fragmenting region has little dependence or almost constant behavior as can be seen from figures 1-3.

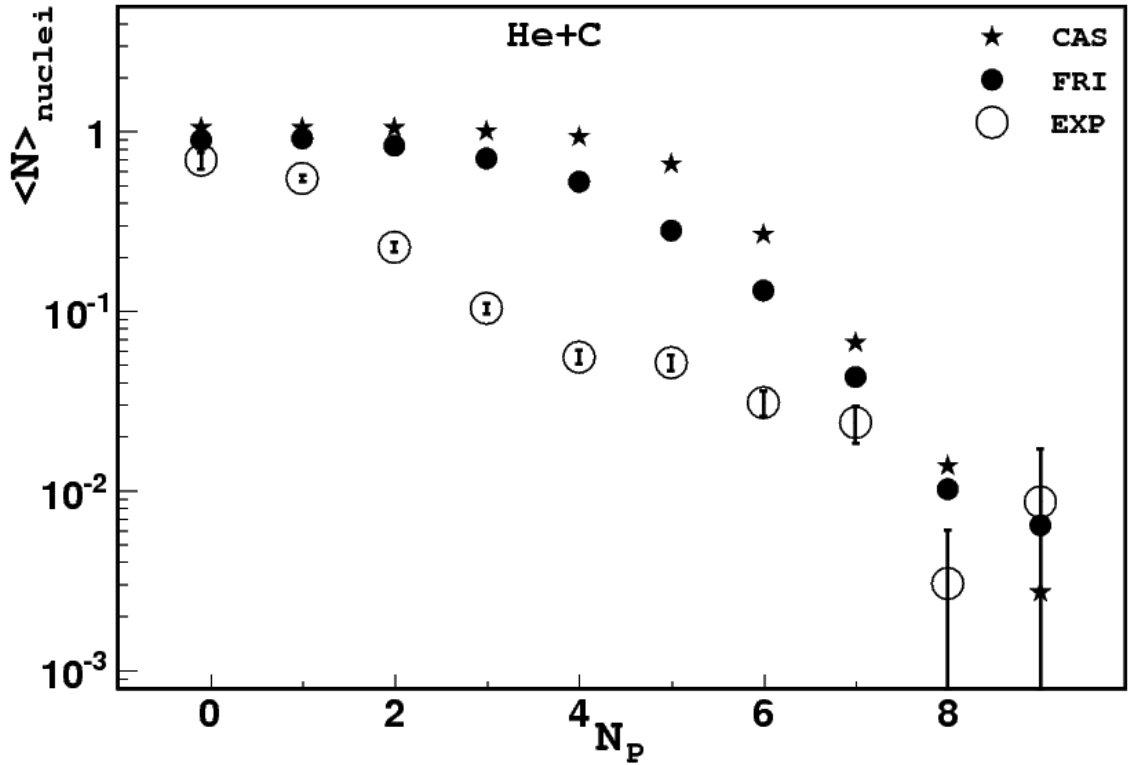

Figure 3. Average multiplicity of light nuclei as function of centrality ($N_p$) in *HeC* - interactions. Stars, solid circles and open circles represent Cascade (CAS), Fritiof (FRI) and data (EXP) respectively

The average multiplicity of the light nuclei ($<N>_{nuclei}$) in $^{12}CC$-interactions at 4.2A GeV/c as a function of centrality is presented in Figure 4, statistical uncertainties are incorporated only. Now the projectile mass and energy is much greater than the above interactions, which results in an increase in multiplicity and better identification of light nuclei. Experimental results are described qualitatively by dividing them into different regions as shown in fig. 4. The measurements are compared with the models' predictions also. Both the models overestimate the average multiplicity in the region ($N_p \leq 2$) by a small amount. In experimental data maximum number of light nuclei is found in peripheral collisions ($N_p \leq 2$) where the impact parameter is large. In this region the Cascade model measures the average multiplicity 1.46 times and the

Fritiof measures 1.3 times of the experimental data. As the impact parameter decreases and collisions become more central ( $N_p \leq 7$) the projectile starts fragmenting into hadrons rather than light nuclei so the average multiplicity of light nuclei decreases linearly with a slope of -0.17 ± 0.005. These nuclei are mostly projectile fragments and this area ($N_p \leq 7$) can be considered as projectile fragmentation or semi-central region as discussed above for HeC- interactions. In this area Cascade model over estimates the multiplicity and Fritiof predictions are nearly in accord with experimental data and both have the behavior similar to that of the experimental data. The deviation of models from experimental data becomes lesser than in the peripheral region, Cascade measurements are about 1.12 times and the Fritiof measurements are about same or less than the experimental observations. Furthermore, when the interactions are more central ($8 \leq N_p \leq 12$) the $<N>_{nuclei}$ decreases more slowly with a slope of -0.094 ± 0.005. It can be expected that the light nuclei from target fragments are more contributing in this region than the above region, which changes the slope of decrease. This central region can be considered as some mixture of projectile and target fragmentation. In this region models' measurements are low as compared to the experimental results. The different behavior of multiplicity of light nuclei as a function of centrality emitting from projectile and target are also discussed in Ref. 6 and the yields of Deuteron and Triton from target spectator area are discussed in Ref. 7. So the studies of light nuclei production in other experiments also give some clues to distinguish projectile and target regions. In the most central collisions ($N_p>12$), in contrast with models, a minor increase in the average multiplicity of light nuclei is observed. This increase in average multiplicity and the decrease in the slope of central region indicate a new source of light nuclei production other than the fragmentation of projectile and target. It is suggested that the new source may be the coalescence mechanism, because in the most central events the collisions are head on and more possibly, the projectile and the target disintegrate into hadrons instead of nuclei. A dense medium is expected in the most central events due to maximum number of participant, in which the nucleons within the same phase space coalesce to make nuclei. Light nuclei production via coalescence mechanisms is predicted in experiment E864 [2] for 10% most central events in Au+Pt (Pb) (for heavy ion collisions) interactions at 10.6 A GeV/c. In our study we find some direct and sharp signatures of nuclear coalescence effect for $^{12}$CC (light ion collisions) interactions at 4.2A Gev/c. This information is necessary for theoretical models to describe the dynamics of the coalescence effect at high energy hadron-nuclear and nuclear-nuclear interaction.

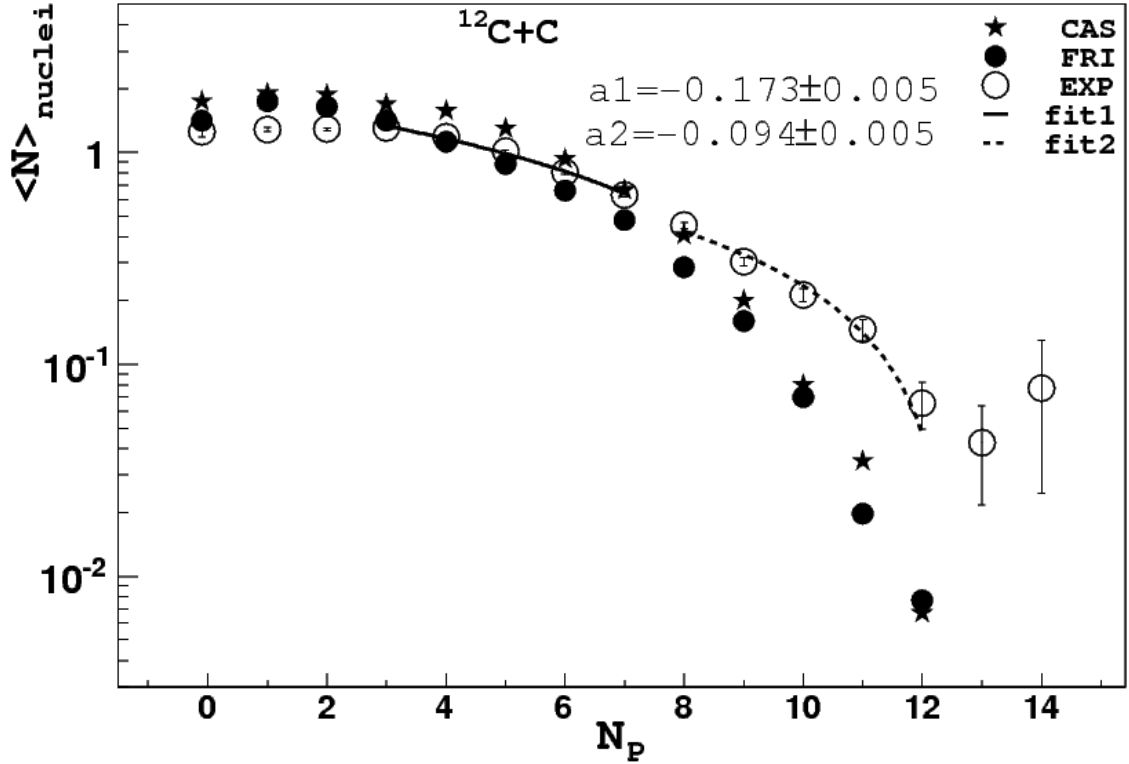

Figure 4. Average multiplicity of light nuclei as function of centrality ($N_p$) in $^{12}CC$ - interactions. Stars, solid circles and open circles represent Cascade (CAS), Fritiof (FRI) and data (EXP) respectively

**Summary**

In summary, analyses of experimental data for the average multiplicity of light nuclei as a function of centrality in *pC*, *dC*, *HeC* and $^{12}CC$ -interaction and their comparison with models are presented. With the increase in the mass of projectile average multiplicity of nuclei increases and diversity between models and experiment decreases. A Systematic change in behavior of projectile can be seen clearly from figs. 1-4 (*pC* to $^{12}CC$). In $^{12}CC$ interactions (fig. 4) we identified four regions ($N_p \geq 2$, $2 < N_p \leq 7$, $8 \leq N_p \geq 11$ and $N_p > 12$), in regions ($N_p > 7$) average multiplicity decreases with less slope and at ($N_p \geq 12$) a minor increase in multiplicity is seen which, indicates a mechanism of light nuclei formation other than the fragmentation of the colliding nuclei, which could possibly be the nuclear coalescence effect.

**Acknowledgement**

We would like to thank V. V. Uzhinskii for providing, and helping to use, modified version of Fritiof model.